%% file: microswim.tex
\documentclass[notitlepage,onecolumn,groupedadress,amsfonts,10pt,tightenlines,floatfix]{revtex4-1}

\usepackage[utf8]{inputenc}
\usepackage{graphicx}
\usepackage{amsmath,amssymb}
\usepackage{upgreek}

\usepackage{amsmath,amssymb,amsfonts,graphicx,bm,bbm,mathtools,sidecap}

\let\citenum\cite 
\let\varpsi\psi
\let\updelta\delta
\let\vec\mathbf
\mathtoolsset{showonlyrefs=true}
\sidecaptionvpos{figure}{c}
\usepackage{xcolor}
\usepackage{float}
\usepackage{nicefrac}
\usepackage[activate={true,nocompatibility},final,tracking=true,kerning=true,spacing=false,factor=1100,stretch=0,shrink=50]{microtype}
\usepackage{makeidx}

\begin{document}
\vspace*{-3em}
\title{Stationary shapes of deformable particles moving at low Reynolds numbers}
\author{Horst-Holger Boltz}
\email[]{horst-holger.boltz@udo.edu}

\author{Jan Kierfeld}
\email[]{jan.kierfeld@tu-dortmund.de}

\affiliation{Physics Department, TU Dortmund University, 
44221 Dortmund, Germany}

\date{\today}              
\begin{abstract}Lecture Notes of the Summer School ``Microswimmers -- From
Single Particle Motion to Collective Behaviour'',
organised by the DFG Priority Programme SPP 1726
(Forschungszentrum J{\"{u}}lich, 2015). \end{abstract}
\maketitle        
\vspace*{-3em}
\section{Introduction}

\index{microcapsule}\index{red blood cell}\index{vesicle}\index{droplet}
\index{sedimentation} \index{boundary integral method}
The motion of
deformable micron-sized objects though a viscous fluid represents an important
problem with various applications, for example, for elastic microcapsules
\cite{Barthes-Biesel2011}, red blood cells \cite{Fedosov2014,Freund2014} or
vesicles moving in capillaries, deforming in shear flow, or sedimenting under
gravity \cite{Abreu2014,huang2011}.  Another related system are droplets
moving in a viscous fluid \cite{clift1978bubbles,Stone1994}.  Motion of these
deformable objects can be caused by external body forces, as in sedimentation
under gravity or in a centrifuge, dragging the object through a quiescent
fluid, or, in the absence of driving forces, by putting the object into a
hydrodynamic flow, for example, capillary flow. In the context of
microswimmers, another possibility is self-propulsion of a soft microswimmer,
for example, by fluid flows generated at its surface.  On the micrometer
scale, the hydrodynamic flows involved in the motion of these objects feature
low Reynolds numbers unless the particle velocities become very high.  Many
elastic micron-sized objects, such as capsules, vesicles or red blood cells,
are easily deformable because elasticity only stems from a thin elastic shell
surrounding a liquid core.
The analytical description and the simulation are challenging problems
as the hydrodynamics of the fluid is coupled to the elastic deformation of the
capsule or vesicle. It is important to recognise that this coupling is mutual:
On the one hand, hydrodynamic forces deform a soft capsule, a vesicle, or a
droplet. On the other hand, the deformed capsule, vesicle or droplet changes
the boundary conditions for the fluid flow. As a result of this interplay, the
soft object deforms and takes on characteristic shapes; eventually there are
transitions between different shapes as a function of the driving force or
flow velocity. Such shape changes might have important consequences for
applications or biological function, for example, if we consider red blood
cells or microcapsule containers moving in narrow capillaries. Shapes can
also exhibit additional dynamic features such as tank-treading or tumbling,
as it has been shown experimentally and theoretically for vesicles or elastic
capsules in shear flows \cite{Barthes-Biesel2011,Abreu2014}.
In the following, we investigate stationary shapes of elastic capsules 
sedimenting in an otherwise quiescent incompressible fluid, either by gravity
or by the centrifugal force. Possible shapes and the nature
of dynamic transitions between them are only poorly understood for 
sedimenting capsules.  An elastic capsule is a  closed elastic shell, 
i.e., a  two-dimensional solid, which can support in-plane shear
stresses and inhomogeneous stretching stresses with respect to their 
equilibrium configuration.
This is different from a  fluid vesicle, which is
 governed by bending elasticity only and is bounded
by a two-dimensional fluid surface (lipid membrane)  with
vanishing shear modulus \cite{Seifert1997}.
Whereas the rest shape of vesicles is determined by a
few global parameters such as fixed area and spontaneous curvature 
\cite{Seifert1997}, elastic
capsules can be produced with arbitrary rest shapes, in principle. We focus
on elastic microcapsules with a spherical rest shape.
 We use this problem to introduce a method 
to calculate efficiently axisymmetric stationary
capsules shapes, which iterates between a boundary integral method to solve
the viscous flow problem for given capsule shape and capsule shape equations
to calculate the capsule shape in a given fluid velocity field. The method
does not capture the dynamic evolution of the capsule shape but
converges to its stationary shape.

\section{Prologue -- Cauchy momentum equation}

The unifying concept of both the shape of an elastic capsule and the motion of
a viscous fluid is {\em continuum mechanics}. 
The fundamental equation in point mechanics is Newton's second law
\begin{align}
\frac{\mathrm{d}\vec{P}}{\mathrm{d}t} &= \vec{F} \text{,}
\end{align}
that is the change of the momentum $\vec{P}$ is given by the net force
$\vec{F}$ acting on a mass point. 
 In order to formulate the equivalent equation for a continuous
medium, we consider a volume element $\mathrm{d}V$ with mass $\rho\mathrm{d}V$
moving with velocity $\vec{u}(\vec{r},t)$. We 
 stress the fact that $\vec{u}$ is a field variable:
there is a velocity for every point $\vec{r}$ in the total volume, $\Omega$,
and every time $t$ considered. For the change of momentum of a
specific volume element that moves in time, we have to track its
motion and compute the total time derivative of $\vec{u}(\vec{r}(t),t)$
along the path of its motion \cite{batchelor}, that is with $\partial_t
\vec{r}(t)=\vec{u}$, which is called the {\em material derivative}
\begin{equation}
  \frac{\mathrm{D}(\rho \vec{u})}{\mathrm{D}t} \equiv 
   \frac{ \partial (\rho\vec{u})}{\partial t} 
     + \rho \vec{u} \cdot \nabla (\rho \vec{u})
  \text{.}
\end{equation}
There can be two types of force acting on the volume element,
interfacial forces $\vec{f}_i$ due to the interaction with 
neighbouring  volume elements
acting on the surface and body forces $\vec{b}$ 
acting on the volume,
and we can adapt Newton's law to
\begin{equation}
  \int_{\Omega}\mathrm{d}V\, \frac{\mathrm{D}(\rho \vec{u})}{\mathrm{D}t} =
  \int_{\partial\Omega} \mathrm{d}A \vec{f}_i +
  \int_{\Omega}\mathrm{d}V\,\vec{b} =
  \int_{\Omega}\mathrm{d}V\, \left[ \nabla\cdot\sigma + \vec{b} \right]
  \text{.}
\end{equation}
We wrote $\vec{f}_i=\sigma\cdot\vec{n}$ with the surface normal $\vec{n}$,
where we  call $\sigma$ the {\em stress tensor}\index{stress tensor},
and applied Stokes' theorem in the last equality. 
Because it holds for arbitrary $\Omega$,
 the integrands are equal, which yields the 
{\em Cauchy momentum equation}\index{Cauchy momentum equation}
\begin{equation}
  \frac{\mathrm{D}(\rho \vec{u})}{\mathrm{D}t} = \nabla\cdot\sigma + \vec{b}
  \text{.} \label{eq:cauchy}
\end{equation}
In the following, we  consider  this  equation both for 
volume elements of fluid and for volume elements of the capsule 
membrane. 
For the fluid and for the capsule material there are different 
 {\em constitutive relations}\index{constitutive relation}, which describe  the relation between 
the  stresses $\sigma$ in the medium and  the deformation state or velocity 
field (stress-strain-relation).
The left-hand side of this equation will
be zero both for capsule and fluid, as we consider a stationary capsule shape 
 ($\vec{u}=0$) 
and a stationary, incompressible (or solenoid, if you will) fluid
flow at low Reynolds number. 
In this stationary case, we obtain stress-balance equations
for the fluid and the capsule shape. 
Both equations are coupled:
The fluid stresses  enter the stress-balance for the capsule.   
Moreover, the fluid velocity field is required to be continuous at the 
boundary between capsule and fluid resulting in a {\em no-slip boundary
  condition}\index{no-slip boundary condition} at the capsule surface.

\section{Equilibrium shape of a thin shell} \index{shell theory}

For the parametrisation of the capsule shape, 
we directly exploit the axisymmetry by  working in cylindrical
coordinates. The axis of symmetry is called $z$, the
distance to this axis $r$ and the polar angle $\phi$. 
The shell
is
given by the generatrix $(r(s),z(s))$ which is parametrised in arc-length $s$
(starting at the lower apex with $s=0$ and ending at the upper apex with $s=L$).
The unit tangent  vector $\vec{e}_s$ 
 to the generatrix at $(r(s),z(s))$ defines
an angle $\psi$ via $\vec{e}_s=(\cos \psi, \sin \psi)$, which can be used to
quantify the orientation of a patch of the capsule relative to the axis of
symmetry.
The shape of a thin axisymmetric shell of  thickness $H$
can be derived from 
non-linear shell theory \cite{LibaiSimmonds1998,Pozrikidis2003}. 
A known reference
shape $(r_0(s_0),z_0(s_0))$ (a subscript zero refers to a quantity of the
reference shape;  $s_0 \in [0,L_0]$ is the arc length 
of the reference shape)
is deformed by hydrodynamic forces exerted by the viscous flow.
Each point $(r_0(s_0), z_0(s_0))$ is mapped onto a point 
$(r(s_0), z(s_0))$ in the deformed configuration, 
which induces  meridional and
circumferential stretches, $\lambda_s = ds/ds_0$ and 
$\lambda_\phi = r/r_0$, respectively. 
The arc length element $\mathrm{d}s$ of the deformed
configuration is  $\mathrm{d}s^2 = (r'(s_0)^2 + z'(s_0)^2) \mathrm{d}s_0^2$.
The shape of the deformed  axisymmetric shell
 is given by the solution of a system of first-order
differential equations, henceforth referred to as 
the shape equations\index{shell shape equations}. \nocite{Knoche2013}
These describe stress-balance, i.e., the balance 
 forces and torques  or 
tensions and bending moments acting on a patch of
 the shell, as  shown in Fig.\ \ref{fig:forces_slab}.
Using the notation of Refs.\ \citenum{Knoche2011,Knoche2013} 
these can be written as 

\begin{minipage}{0.6\textwidth}
\mathtoolsset{showonlyrefs=false}
\begin{subequations}
\begin{align}
 s'(s_0) &= \lambda_s \\
 r'(s_0) &= \lambda_s \cos{\varpsi} \\
 z'(s_0) &= \lambda_s \sin{\varpsi} \\
 \varpsi'(s_0) &= \lambda_s \kappa_s \label{eq:se} \\
 \tau'_s(s_0) &= \lambda_s \left( \frac{\tau_\phi - \tau_s}{r} \cos\varpsi
+ \kappa_s q + p_s\right)\label{eq:forceinplane} \\
 m'_s(s_0) &= \lambda_s \left( \frac{m_\phi - m_s}{r} \cos\varpsi
- q + l\right) \label{eq:torque}\\
 q'(s_0) &= \lambda_s \left( -\kappa_s \tau_s - \kappa_\phi \tau_\phi -
\frac{q}{r} \cos\varpsi + p \right)\label{eq:forcenormal} \text{.}
\end{align}
\end{subequations}
\mathtoolsset{showonlyrefs=true}
\end{minipage}
\hspace{0.09\textwidth}
\begin{minipage}{0.3\textwidth}
\begin{figure}[H]
\raggedleft
\resizebox{0.999\textwidth}{!}{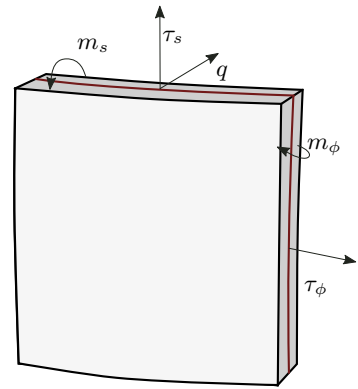}
\caption{Sketch depicting the tensions and moments}
\label{fig:forces_slab}
\end{figure}
\vspace{1.em}
\end{minipage}

The additional 
quantities appearing in these shape equations are defined as follows:
The angle $\varpsi$ is the slope angle between 
the tangent plane to the deformed
shape and the $r$-axis, $\kappa_\phi$ is the
circumferential curvature, $\kappa_s$ the meridional curvature;
$\tau_s$ and $\tau_\phi$ are the 
meridional and circumferential stresses, respectively; $m_s$ and  $m_\phi$
are bending moments; 
$q$ is the transverse shear stress, $p$ the total normal pressure, 
$p_s$ the shear pressure, and $l$ the external stress couple. 
The first equation defines $\lambda_s$, the next three equations
follow from geometry, and the last
three ones express (tangential and normal) force and torque equilibrium.  
All quantities appearing on the right hand side of the shape equations
have to be expressed in terms of the 7 quantities on the left hand side
in order to close the equations.

The curvatures and
circumferential strains  are known from  geometrical relations
$
 \kappa_\phi = \frac{\sin\varpsi}{r}$, 
$\kappa_s = \frac{\mathrm{d}\psi}{\mathrm{d}s}$ and
 $\lambda_\phi = \frac{r}{r_0}$.
The elastic tensions $\tau_s$ and $\tau_\phi$ and bending moments 
$m_s$ and  $m_\phi$, which define the elastic stresses
in the shell material  are related to the  strains  and curvatures
by the material-specific {\em constitutive 
relations}\index{elastic constitutive relation}, 
which relate the stress tensor in a material to the strain tensor 
and involve the elastic constants of the material. 
Such constitutive relations often derive from an elastic energy 
functional, such that the stress tensor components are
 the first variation of the 
energy functional with respect to the corresponding strain tensor
components \cite{LL7}. 
We will derive the general relation between stress tensor and 
 elastic tensions and bending moments of the shell in the following 
section. 

Below, we will focus on  Hookean 
capsules, where the constitutive relations derive from an 
 elastic energy which is quadratic in stretching strains and 
bending strains. This leads to  \cite{LibaiSimmonds1998,Knoche2013}
\begin{align}
\tau_{s}  &=   \frac{Y_{\rm 2D}}{1-\nu_{\rm 2D}^2} \frac{1}{\lambda_\phi}
  \left[ (\lambda_s-1) + \nu_{\rm 2D} (\lambda_{\phi}-1)\right]\\
m_{s}  &=   E_B \frac{1}{\lambda_\phi}
  \left[ (\lambda_s\kappa_s -\kappa_{s0}) + \nu_{\rm 2D} (\lambda_\phi 
  \kappa_\phi -\kappa_{\phi 0})\right]
\end{align}
where $Y_{\rm 2D}$ is the surface Young modulus (which, for isotropic shells, is 
related to the bulk Young modulus by $Y_{\rm 2D} = Y_{\rm 3D} H$),
$\nu_{\rm 2D}$ is the surface Poisson ratio, and $E_B$ is the bending 
modulus  of the shell ($E_B \propto Y_{\rm 2D}H^2$ for isotropic shells);
$\kappa_{s0}$ and $\kappa_{\phi 0}$ are the curvatures of the reference
shape. 
The constitutive relations for $\tau_\phi$ and $m_\phi$ 
 are obtained by interchanging all indices $s$  and $\phi$.

The normal pressure  
\begin{align}
 p&=p_0+p_n + p_{\text{ext}}, 
\label{eq:p}
\end{align}
the shear-pressure $p_s$, and the stress couple\footnote{The fluid inside the
  capsule is assumed to be at rest.} $l=p_s H/2$ are given externally by
hydrodynamic and external forces. The static pressure $p_0$ is the
pressure difference between the interior and exterior liquids. For the case of
a capsule that is filled with an incompressible fluid its value is fixed by
demanding a fixed enclosed volume.

The pressures $p_n$ and $p_s$ are the normal and tangential forces per area
which are generated by the surrounding fluid. For the latter it would usually
be more natural to give express stresses in terms of their radial and axial
contributions.  The vector\footnote{The subscript $H$ indicates that this only
  incorporates the hydrodynamic contributions. We do not consider cases with a
  static shear-pressure but, as is apparent from eq.\ (\ref{eq:p}), there are
  static contributions to the normal pressure.}
\begin{align}
\vec{p}_H &= p_n \vec{n} - p_s \vec{e}_s,
\end{align} 
where $\vec{n}$ 
and $\vec{e}_s$ are the  normal and tangent unit vectors to the generatrix,
$\vec{n} = -\cos\psi \vec{e}_z + \sin\psi \vec{e}_r$ and  
$\vec{e}_s =  \sin\psi \vec{e}_z + \cos\psi \vec{e}_r$, 
equals  the  hydrodynamic surface force density vector 
\begin{align}
\vec f = f_z \vec e_z + f_r\vec e_r \text{,}
\end{align}
which will be calculated below.
Re-decomposing $\vec f$ into its normal and
tangential components $p_n$ and $p_s$ we find
\begin{align}
p_n &= f_r\sin\varpsi - f_z\cos\varpsi \quad \text{and}\quad
p_s = -f_r\cos\varpsi -  f_z\sin\varpsi \text{.}
\end{align}

\subsection{Derivation of the shape equations}

In the following, we present a compact 
derivation of the shape equations (\ref{eq:forceinplane}), (\ref{eq:torque})
and (\ref{eq:forcenormal}) describing force and torque balance. 
We refer the reader to the
literature \cite{LibaiSimmonds1998,Pozrikidis2003,Knoche2011} for 
more elaborate derivations. From the Cauchy momentum
equation we gather that the equilibrium is given by\footnote{External forces
  are easily added and actually necessary for the existence of non-trivial
  solutions.}  $\nabla\cdot\sigma = 0$. We will not use Cartesian
coordinates here, but curvilinear coordinates that are better suited to the
capsule geometry. In the vicinity of the capsule we can parametrise space by
the set of coordinates $n,s,\phi$ and
\begin{align}
  \vec{r}(n,s,\phi)&=
    \left(\begin{array}{c} r(s) \cos\phi \\ r(s)\sin\phi \\
      z(s) \end{array}\right)
     +n \left(\begin{array}{c} \sin\psi \cos\phi \\
        \sin\psi\sin\phi \\ -\cos\psi \end{array}\right)
\end{align}
which corresponds to a local tripod of orthogonal vectors\footnote{
$\tilde{\vec{n}}$ is a three-dimensional normal vector of the axisymmetric
surface, whereas 
$\vec{n}$ is a two-dimensional normal vector to the 
 generatrix $(r(s),z(s))$.}
\begin{align}
  \tilde{\vec{n}}&=\left(\begin{array}{c} \sin\psi \cos\phi \\
      \sin\psi\sin\phi \\ -\cos\psi \end{array}\right)\text{,}~~
  \tilde{\vec{s}}= (1+n\psi')\left(\begin{array}{c} \cos\psi \cos\phi \\
      \cos\psi\sin\phi \\ \sin\psi \end{array}\right)\,\text{and}~~
  \tilde{\boldsymbol{\phi}}= (r+n\sin\psi) \left(\begin{array}{c} \sin\phi \\
      -\cos\phi \\ 0 \end{array}\right) .
\end{align}
Using index notation with $i,j,k \in {n,s,\phi}$, the tripod vectors
can be written as  $\tilde{\vec{i}}=\partial_i \vec{r}$.
We want to derive equations for the elasticity of a {\em thin shell}, which we
treat as an effectively two-dimensional surface. This is done by
restricting stresses to in-plane stresses and integrating over the normal
direction (the $n$-direction, $n\in [-H/2,H/2]$ with $H$ being the
thickness). Thus, it is advantageous to define a projector $ \mathsf{P} =
\mathbbm{1} - \tilde{\vec{n}}\tilde{\vec{n}}$  
that projects the stress onto the subspace of
in-plane stresses. We then have to compute the tensor derivative
occurring in the in-plane 
stress-balance equation $\nabla\cdot\mathsf{P}\sigma = 0$
in these  curvilinear coordinates. We can use the general result (to be
read with Einstein summation convention)
\begin{align}
  (\nabla\cdot\mathsf{A})_k &= \frac{\partial \mathsf{A}_{ik}}{\partial i} -
  \mathsf{A}_{jk} \Gamma^j_{ii} - \mathsf{A}_{ij}\Gamma^j_{ik}
\end{align}
with $i,j,k \in {n,s,\phi}$ and the Christoffel symbols of the second kind
defined by $ \frac{\partial \tilde{\vec{i}}}{\partial j} = \Gamma^{k}_{ij}
\tilde{\vec{k}}$.
The Christoffel symbols of the second kind are symmetric
in the lower indices. For the non-vanishing
symbols (at $n=0$, i.e.\ on the surface, where we want to evaluate the stress)
we find 
\begin{align*}
& \Gamma^s_{ns} = \kappa_s \text{,}\,
 \Gamma^\phi_{n\phi} =\kappa_\phi \text{,}\,
 \Gamma^n_{ss} = - \kappa_s \text{,}\,
 \Gamma^\phi_{s\phi} = \frac{\cos\psi}{r} \text{,}\,
 \Gamma^n_{\phi\phi} = -r\sin\psi\, \text{ and }\,
 \Gamma^s_{\phi\phi} = -r\cos\psi  \text{.}
\end{align*}
Here, we made use of the geometrical relations $\psi'=\kappa_s$ and $\sin\psi =
r\kappa_\phi$. From the given axisymmetry, we  infer the following form of the
projected stress tensor
\begin{align}
  \mathsf{P}\sigma &= \left( \begin{array}{ccc} 0 & 0 & 0 \\ \sigma_{sn} &
      \sigma_{ss} & 0 \\ 0 & 0 &\sigma_{\phi\phi} \end{array} \right)
\end{align} 
and computing the divergence (at $n=0$) finally yields (remember
$\partial_s r= \cos\psi$) 
\begin{align}
  \nabla\cdot\mathsf{P}\sigma &= 
   \left(\frac{1}{r} \frac{\partial
      (r\sigma_{sn})}{\partial s} - \kappa_s \sigma_{ss} - \sigma_{\phi\phi}
    \kappa_{\phi}\right)\tilde{\vec{n}} + \left(\frac{1}{r}\frac{\partial
      (r\sigma_{ss})}{\partial s} + \kappa_s \sigma_{sn} - \sigma_{\phi\phi}
    \frac{\cos\psi}{r} \right) \tilde{\vec{s}} \text{.}
\end{align}
If we integrate over the small thickness and introduce (the sign of $q$ might
appear random and is such that the definition of $q$ agrees with the
literature)
\begin{align}
\int_{-H/2}^{H/2} \mathrm{d}n\, \sigma_{ss} &= \tau_s \text{,}\quad
\int_{-H/2}^{H/2} \mathrm{d}n\, \sigma_{\phi\phi} = \tau_\phi \quad \text{and}\quad
\int_{-H/2}^{H/2} \mathrm{d}n\, \sigma_{sn} = -q \text{.}
\end{align}
The in-plane 
stress-balance $\nabla\cdot\mathsf{P}\sigma = 0$ then 
 establishes the  shape equations (\ref{eq:forceinplane}) and
 (\ref{eq:forcenormal}). 

The third equation (\ref{eq:torque}) is
obtained  from considering the acting torque. A shell of finite thickness is
able to sustain finite interfacial 
bending moments $\vec{m}$, which, in terms of basic
physics, means that there is an additional contribution to the torque
balance. Arguing along the same lines as we did for the Cauchy momentum
equation but starting from Newton's second law for rotational
motion,
\begin{align}
\frac{\mathrm{d}L}{\mathrm{d}t} &= \vec{M} \text{,}
\end{align}
yields (in equilibrium)
\begin{align}
  0 &= \int_{\Omega}\mathrm{d}V (\vec{r}'\times \vec{b}) 
     + \int_{\partial\Omega}\mathrm{d}A
      (\vec{r}'\times\sigma \vec{n}_A + \vec{m} )
\end{align}
where $\vec{r}'$ is the difference vector to the centre of mass of the
patch. Now we introduce the {\em bending moment tensor} (also couple stress
tensor) $\mathsf{M}$ by $ \vec{m} = \mathsf{M} \vec{n}_A $
($\vec{n}_A$ is the three-dimensional normal to the surface $A$)
 and the auxiliary
tensor $\mathsf{D}$ by $ \mathsf{D}\cdot\vec{n}_A = \vec{r}'\times
\sigma\cdot\vec{n}_A$;
 after once again applying the divergence theorem we find, in absence
of body forces, $\nabla\cdot\left(\mathsf{D}+\mathsf{M}\right) = 0$.
We are interested in the axisymmetric case $\vec{n}_A=\tilde{\vec{n}}$ 
and (after
projection to in-plane stresses) see that $\sigma\tilde{\vec{n}} = \sigma_{sn}
\tilde{\vec{s}}$. We consider a small patch such that we can write $\vec{r'}=
s\tilde{\vec{s}}+n\tilde{\vec{n}}+\phi\tilde{\vec{\phi}}$ and find after
projection to in-plane-torques and using that the internal bending moments act
along the directions of principal curvature, i.e.
\begin{align}
\mathsf{P}\mathsf{M} &= \left( 
\begin{array}{ccc} 0 & 0 & 0 \\ 0 & M_{ss} & 0 \\
 0 & 0 & M_{\phi \phi} \end{array}\right) 
\text{,}
\end{align}
for $n=0$
\begin{align}
0 &= \frac{1}{r}\frac{\partial rM_{ss}}{\partial s} 
   + \sigma_{sn} -\frac{\cos\psi}{r} M_{\phi\phi}
\end{align}
which  gives  equation (\ref{eq:torque}) after renaming
\begin{align}
\int_{-H/2}^{H/2} \mathrm{d}n\, M_{ss} &= m_s \quad\text{and}\quad
\int_{-H/2}^{H/2} \mathrm{d}n\, M_{\phi\phi} = m_\phi \text{.}
\end{align}

\subsection{Solution of the shape equations.}

The boundary conditions for a shape that is closed and has no kinks at its
poles are
\begin{align}
  &r(0) = r(L_0) = \psi(0) = \pi - \psi(L_0) = 0,
\label{eq:rpsi0}
\end{align}
and we can always choose\footnote{In the presence of gravity there is no
  translational symmetry along the axial direction, but shifting the capsule
  as a whole just adds a constant to the hydrostatic pressure $-g\Delta \rho z
  \rightarrow -g\Delta \rho (z+z_0)$ which is absorbed into the static
  pressure $p_0$.} $z(0)=0$.  If hydrodynamic drag and gravitational pull
cancel each other in a stationary state, there is no remaining point force at
the poles needed to ensure equilibrium and, thus,
\begin{align}
 q(0)& = q(L_0) = 0 \text{.}
\label{eq:q0}
\end{align}
The shape equations have (removable) singularities at both poles; therefore, a
numerical solution has to start at both poles requiring $12$ 
boundary conditions ($r,z, \psi, \tau, m, q$ at both poles)
out of which we know
$7$ (by eqs.\ \eqref{eq:rpsi0} and \eqref{eq:q0} and $z(0)=0$). 
The $5$ remaining parameters  can be determined by a shooting method 
requiring
that the solutions starting at $s_0=0$ and  $s_0=L_0$ have to
match continuously  in the middle, 
which gives $6$  matching conditions ($r,z,\psi, \tau, m,
q$). This gives an over-determined non-linear set of equations which we solve
iteratively using linearizations. 
However, as in the static
case \cite{Knoche2011}, the existence of a solution to the resulting system of
linear equations (the matching conditions) is ensured by the existence of a
first integral (see below) of the shape equation. In principle,
this first integral could be
used to cancel out the matching condition for one parameter (e.g.\ $q$),
such that  the system
is not genuinely over-determined. We found the
approach using an over-determined system to be better numerically 
tractable, where  we ultimately used a multiple 
shooting method\index{shooting method}
including several matching points between the poles.

Using these boundary conditions, it is straightforward to see that the
shape equations 
 do not allow for a solution whose shape is the reference shape, 
unless there are no external loads ($p_s=p=l=0$).

\subsection{First Integral of the shape equations}\label{app:first}

We make the
following Ansatz (cf.\ eqs.\ (17), (22) in Ref.\ \citenum{Knoche2011}) 
for a first integral of the shape equations
\begin{align}
 U(s) &=  2\pi r \cos\psi q + 2\pi r\sin\psi \tau_s
   + X =\text{const.}
\label{eq:Us}
\end{align}
We are also assuming that the pressure $p$ and the shear pressure $p_s$ can be
written as functions of  the arc length $s$ only. The calculation
is rather straightforward, we differentiate and get
\begin{align}
 0 &= U'(s) \nonumber \\
 &= 2\pi \cos^2\psi q - 2\pi r \sin\psi \kappa_s q 
 + 2\pi r\cos\psi (-\kappa_s \tau_s - \kappa_\phi \tau_\phi 
     - \cos\psi \frac{q}{r} +p) +2\pi \cos\psi \sin\psi \tau_s \nonumber\\
 &~+ 2\pi r
  \cos\psi\kappa_s\tau_s+2\pi r\sin\psi ( \cos\psi
  \frac{\tau_\phi-\tau_s}{r} + \kappa_s q + p_s)   + X' 
\\
&= 2 \pi r p \cos \psi + 2\pi r p_s \sin\psi + X' = 0.
\end{align}
In the second to last step most terms cancel each other out.
 Thus, we arrive at 
an ordinary differential equation for $X$, which we can integrate 
to find 
\begin{align}
 X(s)  &= -2\pi \int_0^s \mathrm{d}x\, r(p\cos\psi+p_s\sin\psi).
\label{eq:appX}
\end{align}
Inspecting the behaviour at $s=0$ we deduce $U(s)=U(0)=0$,  
which implies $U(L)=0$ and, according to (\ref{eq:Us}), $X(L)=0$.  
The physical interpretation of $X(L)=0$ in (\ref{eq:appX}) 
 is that the capsule has to
be in global force balance. By symmetry there can be no net force in radial
direction, but the external forces can lead to net force in axial direction.
The quantity $X$ contains the contribution to the net force in $z$-direction
and thus a shape with the desired features (namely $q=0$ at the apexes) must
have $X(L)=0$ and, thus, be in global force balance.

\section{Low Reynolds-number Hydrodynamics}

We want to calculate the flow field of a viscous incompressible fluid around
an axisymmetric capsule of given fixed shape at low Reynolds numbers.  We
chose to separate the problems, which allows us to state that for the
calculation of the flow field the deformability of the capsule is not
relevant, and the capsule can be viewed as a general immersed body of
revolution $\mathfrak{B}$.  For the calculation of the capsule shape, which is
addressed in the following section, and for the determination of its
sedimenting velocity, we only need to calculate the surface forces onto the
capsule which are generated by the fluid flow.
 
 \subsection{Stokes equation}\index{Stokes equation}
 
 We start with the fundamental notion that the mass of the fluid should be
 conserved under its flow $\vec{u}$ giving rise to the {\em continuity
   equation}
 \begin{align}
 \frac{\partial \rho}{\partial t} &= - \nabla\cdot(\rho \vec{u}) 
 \end{align}
 with $\rho$ being the local mass density. Furthermore, we only consider
 incompressible fluids, thus, the density of  a fluid volume element 
 cannot change under its motion due to the flow or
 \begin{align}
 0= \frac{D \rho}{D t} &= \frac{\partial \rho}{\partial t} 
   + \vec{u}\cdot\nabla\rho \text{.}
 \end{align}
We combine these two equations and  find with some help from vector calculus
\begin{align}
\nabla\cdot \vec{u} &= 0 \text{,} 
\label{eq:incomp}
\end{align}
which is commonly referred to as the continuity equation for an incompressible
flow. Using this equation in the
Cauchy momentum equation, eq.\ \eqref{eq:cauchy}, 
 for a {\em stationary flow} with $\partial_t \vec{u}=0$
gives\footnote{We omit the
  external force as they do not change any of the following in a non-trivial
  manner.}
\begin{align}
\rho \vec{u}\cdot \nabla\vec{u} &= \nabla\cdot\sigma \text{.} 
\label{eq:tempeq}
\end{align}
For further progress, we need  the {\em constitutive
  relation}\index{fluid constitutive relation} for the
liquid. Demanding Galilean invariance of $\sigma$, we see that $\sigma$ can
only depend on spatial derivatives of the velocity and from conservation of
angular momentum we know that $\sigma$ is symmetric\footnote{This is sometimes
  referred to as Cauchy's second law of motion (the first one being the
  momentum equation).}. Based on experimental evidence we further demand that
there are no shear stresses in a quiescent fluid, that $\sigma$ is isotropic
and that stresses grow linear with the velocity, which allows us to write
\begin{align}
\sigma &= -p \mathbbm{1}
  +\mu \left(\nabla\vec{u} + (\nabla\vec{u})^T \right) \text{.} 
\label{eq:sigma}
\end{align}
or, in Cartesian coordinates, 
$
 \sigma_{ij} = -p \updelta_{ij} + 2 \mu e_{ij} \text{,}
$
with the pressure $p$, the viscosity $\mu$ and the 
{\em rate of deformation}\index{rate of deformation tensor}
(or rate of strain) tensor for the flow velocity $\vec{u}$
\begin{align}
 e_{ij} &= \frac{1}{2} \left( \frac{\partial u_i}{\partial x_j} 
      + \frac{\partial u_j}{\partial x_i} \right) \text{.}
\end{align}
A liquid for which our assumptions hold is called a {\em Newtonian
  fluid}\index{Newtonian fluid}. 
The constitutive relation together with eq.\ \eqref{eq:tempeq}
(stationary Cauchy  momentum equation for an incompressible fluid) 
give rise to the (stationary) {\em Navier-Stokes
  equation}\index{Navier-Stokes equation}. Rescaling  velocities,
lengths and stresses by their respective typical scales $u$, $R$, $\mu v/L$
(as inferred from the constitutive relation)
in  eq.\ \eqref{eq:tempeq} we obtain 
\begin{align}
\mathrm{Re}\, \bar{\vec{u}}\cdot\bar{\nabla}\bar{\vec{u}} &= 
\bar{\nabla}\cdot\bar{\sigma}
\end{align} 
for the corresponding dimensionless  quantities $\bar{\vec{u}}$ and 
$\bar{\sigma}$
with the {\em Reynolds number}\index{Reynolds number}
\begin{align}
\mathrm{Re} &= \frac{\rho u L}{\mu} \text{.}
\end{align}
We can neglect the so-called advective term on the left hand-side of our
equation of motion, if the Reynolds number is sufficiently low, that is for
small, slowly moving particles in a medium of high viscosity.
All in all, in the limit of small Reynolds numbers
in a stationary fluid in the absence of  external body forces, 
the stress tensor $\sigma$ is given by
the {\em stationary Stokes equation} \cite{HappelBrenner1983}
\begin{align}
 \nabla\cdot \sigma &= 0 \text{.}
\label{eq:Stokes}
\end{align}

\subsection{Lorentz' reciprocal theorem}\index{Lorentz' reciprocal theorem}

As a preliminary for the following, we derive a relation between two solutions
of the Stokes equation, commonly known as Lorentz' reciprocal
theorem\footnote{This is an application of Green's second identity of vector
  calculus.}. Suppose we have two velocity fields
$\tilde{\vec{u}}$, $\widehat{\vec{u}}$ with corresponding stress tensors
$\tilde{\sigma}$, $\widehat{\sigma}$ both of which solve the 
Stokes equation \eqref{eq:Stokes} with the constitutive relation
\eqref{eq:sigma}. Now, for reasons that will become apparent instantly, we look
at the following rather odd expression
\begin{align}
  \tilde{\vec{u}} \cdot \left( \nabla \cdot \widehat{\sigma}\right) &=
  \nabla\cdot \left( \tilde{\vec{u}} \widehat{\sigma} \right) - \left(
    \widehat{\sigma} \nabla\right) \tilde{\vec{u}} = \frac{\partial}{\partial
    x_j} \left( \tilde{u}_i \widehat{\sigma}_{ij} \right) - \mu \left(
    \frac{\partial \widehat{u}_i}{\partial x_j} + \frac{\partial
      \widehat{u}_j}{\partial x_i} \right) \frac{\partial
    \tilde{u}_i}{\partial x_j} \text{.} \label{eq:rec1}
\end{align}
where we inserted \eqref{eq:sigma} and exploited that the pressure term
vanishes due to the continuity equation. Subtracting \eqref{eq:rec1} from its
counterpart with tildes and hats interchanged we find that the terms involving
the viscosity cancel out yielding
\begin{align}
  \widehat{\vec{u}} \cdot \left( \nabla \cdot
    \tilde{\sigma}\right)-\tilde{\vec{u}} \cdot \left( \nabla \cdot
    \widehat{\sigma}\right) &= \nabla\cdot \left( \widehat{\vec{u}}
    \tilde{\sigma} \right)- \nabla\cdot \left( \tilde{\vec{u}}
    \widehat{\sigma} \right) \text{.}
\end{align}
Given that we assumed $\tilde{\sigma}$,$\widehat{\sigma}$ being solutions of
eq.\ \eqref{eq:Stokes} we know that the left-hand side of the last equation is
zero and we find the {\em reciprocal identity}
\begin{align}
  0 &=\nabla\cdot \left( \widehat{\vec{u}} \tilde{\sigma} - \tilde{\vec{u}}
    \widehat{\sigma} \right) \text{.} \label{eq:Lorentz}
\end{align}

\subsection{Solution of the Stokes equation (in an axisymmetric domain)}

In the rest frame of the sedimenting axisymmetric capsule and 
with   a ``no-slip'' condition at the capsule surface $\partial\mathfrak{B}$, 
we are looking for axisymmetric solutions that have a given
flow velocity $\vec{u}^\infty$ at infinity and vanishing velocity on the
capsule boundary. 
In the lab frame, $-\vec{u}^\infty$ is  the sedimenting 
velocity of the capsule in the stationary state.
Therefore, $\vec{u}^\infty$ has to be determined 
by balancing the total gravitational pulling force 
and the total hydrodynamic drag force on the capsule. 
For the calculation of the flow field the deformability 
of the capsule is not relevant and the capsule can be viewed 
as a  general  immersed body of revolution $\mathfrak{B}$. 
For the calculation of the total hydrodynamic drag force and 
for the calculation of the capsule shape we need the surface force field 
 $\vec{f}=\sigma\cdot\vec{n}$ generated by the flow, 
where $\vec{n}$ is the
local surface normal. This  is the only property of the 
fluid flow entering the shape equations for the 
capsule and the equation for the sedimenting velocity $|\vec{u}^\infty|$. 
 In the lab frame we are looking for solutions with vanishing pressure and
velocity at infinity. The Green's function for these boundary condition is
the well-known {\em Stokeslet}\index{Stokeslet}\footnote{We are looking for solutions of 
the Stokes equation with an external point force, 
\begin{align}
0&=\nabla \cdot \sigma + \vec{F}_p \updelta(\vec{x})
   = \nabla p - \mu \Delta \vec{u}+ \vec{F}_p \updelta(\vec{x}).
\end{align}
Taking the divergence  and using  $\nabla\cdot \vec{u}=0$
we obtain 
$\Delta p = -\vec{F}_p \cdot \nabla \delta(\vec{x})$.
This equation is solved (analogously to electrostatics) by 
\begin{align}
p &= \vec{F}_p\cdot  \nabla \frac{1}{4\pi x} = 
   -\frac{\vec{F}_p\cdot \vec{x}}{4\pi x^3}.
\end{align}
Using this in the Stokes equation with point force, 
one obtains $\mu \vec{u} = -\vec{F}_p(\nabla\nabla - \Delta \mathbbm{1}) 
   h$, where $h= -x/(8\pi)$ is the solution of
 $\Delta \Delta h = \delta(\vec{x})$, i.e., $\Delta h = -1/(4\pi x)$. 
This leads to the Stokeslet and
 Stresslet. 
 } (also called Oseen-Burgers tensor), that is the fluid
velocity $\vec{u}$ at $\vec{y}$ due to a point-force $-\vec{F}_p$ at $\vec{x}$
\begin{align}
 \vec{u}(\vec{y}) &= -\frac{1}{8\pi \mu} \mathsf{G}(\vec{y}-\vec{x}) \cdot
\vec{F}_p
\label{eq:uGF}
\end{align}
with the Stokeslet $\mathsf{G}$ whose elements are in Cartesian coordinates
($x=\lvert \vec{x}\rvert$)
\begin{align}
 \mathsf{G}_{ij}(\vec{x}) &= \frac{\delta_{ij}}{x}+\frac{x_i x_j}{x^3} 
\text{.}
\end{align}
The corresponding stress tensor is given by
\begin{align}
\sigma_{ij}(\vec{y}) &= 
   \frac{1}{8\pi} \mathsf{T}(\vec{y}-\vec{x}) \cdot \vec{F}_p 
\end{align}
with the {\em Stresslet}\index{Stresslet} $\mathsf{T}$ 
whose Cartesian elements are 
\begin{align}
\mathsf{T}_{ijk}(\vec{x}) &= - 6 \frac{x_i x_j x_k}{x^5}  \text{.} 
\end{align}
From the reciprocal theorem \eqref{eq:Lorentz}
we deduce (as $\vec{F}_p$ is a constant) that any
solution of the Stokes equation has to satisfy
\begin{align}
  0&=\frac{\partial}{\partial y_k} \left( \mathsf{G}_{ij}(\vec{y}-\vec{x})
    \sigma_{ik}(\vec{y}) - \mu u_i (\vec{y}) \mathsf{T}_{ijk}
    (\vec{y}-\vec{x}) \right)
\end{align}
or, after integrating over a volume with surface $\partial V$ 
by virtue of Stokes theorem,
\begin{align}
  \iint_{\partial V} \!\mathrm{d}\vec{A}\,\left(
    \mathsf{G}_{ij}(\vec{y}-\vec{x}) \sigma_{ik}(\vec{y}) - \mu u_i (\vec{y})
    \mathsf{T}_{ijk} (\vec{y}-\vec{x})\right) &= 0
\end{align}
For this equation to be valid the volume $V$ must not contain the singularity
at $\vec{x}$. A straightforward way to ensure this is to consider the volume
enclosed by $\partial \mathfrak{B}$ and $\partial B_{\varepsilon} (\vec{x})$,
where the latter is the ball of infinitesimal size $\varepsilon$ around
$\vec{x}$. Separating the surface integral this leads to
\begin{align}
  &\iint_{\partial \mathfrak{B}} \!\mathrm{d}\vec{A}\,\left(
    \mathsf{G}_{ij}(\vec{y}-\vec{x}) \sigma_{ik}(\vec{y}) - \mu u_i (\vec{y})
    \mathsf{T}_{ijk} (\vec{y}-\vec{x})\right) \\
  &= - \iint_{\partial
    B_{\varepsilon} (\vec{x})} \!\mathrm{d}\vec{A}\,\left(
    \mathsf{G}_{ij}(\vec{y}-\vec{x}) \sigma_{ik}(\vec{y}) - \mu u_i (\vec{y})
    \mathsf{T}_{ijk} (\vec{y}-\vec{x})\right)
\end{align}
In the limit $\varepsilon\rightarrow 0$ we can simplify the right-hand side
with $\vec{z}=\vec{y}-\vec{x}$ and
$\mathrm{d}\vec{A}=\vec{n}\mathrm{d}A=\varepsilon^{-1}\vec{z} \varepsilon^2
\mathrm{d}\Omega$ ($\Omega$ being the solid angle)
\begin{align}
  &\iint_{\partial B_{\varepsilon} (\vec{y})} \!\!\!\!\!\!\!\!\!\!\mathrm{d}\vec{A}\,\left(
    \mathsf{G}_{ij}(\vec{y}-\vec{x}) \sigma_{ik}(\vec{x}) - \mu u_i (\vec{y})
    \mathsf{T}_{ijk} (\vec{y}-\vec{x})\right) +\mathcal{O}(\varepsilon) 
  \\
  &=
  \iint_{\partial B_{\varepsilon} (\vec{y})} \!\!\!\!\!\!\!\!\!\!\mathrm{d}\Omega\,z_k\left(
    \left[\delta_{ij}+\frac{z_i z_j}{\varepsilon^2} \right]
    \sigma_{ik}(\vec{x}) + \mu u_i (\vec{x}) \left[ 6 \frac{z_i z_j
        z_k}{\varepsilon^4} \right]\right) =\iint_{\partial B_{\varepsilon}
    (\vec{y})} \!\!\!\!\!\!\!\!\!\!\mathrm{d}\Omega\,z_k\left( \mu u_i (\vec{x}) \left[ 6
      \frac{z_i z_j z_k}{\varepsilon^4} \right]\right) \\&=\frac{6\mu
    u_i(\vec{x})}{\varepsilon^2}\iint_{\partial B_{\varepsilon} (\vec{y})}
  \!\!\!\!\!\!\!\!\!\!\mathrm{d}\Omega\,\left( z_i z_j\right)= \frac{6\mu
    u_i(\vec{x})}{\varepsilon^2} 2\pi \varepsilon^2 \delta_{ij} \int_{-1}^{1}
  \!\mathrm{d} ( \cos\theta ) \cos^2\theta = 8 \pi \mu u_j(\vec{x})
\end{align}
and thus gather the 
{\em boundary integral formula}\index{boundary integral formula},
 which we express as a
function of the acting surface forces $\vec{f}=\sigma\cdot \vec{n}$ including a
constant velocity accounting for the centre of mass motion of the capsule
\begin{align}
  u_j(\vec{x}) - u_j^{\infty}(\vec{x}) &= 
  -\frac{1}{8\pi\mu} \iint_{\partial
    \mathfrak{B}} \!\mathrm{d}A\, \mathsf{G}_{ij}(\vec{y}-\vec{x})
  f_i(\vec{y}) + 
  \iint_{\partial \mathfrak{B}} \!\mathrm{d}\vec{A}\,
  \frac{1}{8\pi} u_i (\vec{y}) \mathsf{T}_{ijk} (\vec{y}-\vec{x}) \text{.}
\end{align}
The physical interpretation of this equation is that the flow field is, on the
one hand, due to point forces (first term, also called the {\em single layer
  potential})\index{single layer potential} and, on the other hand, 
due to point sources and force dipoles (second
term, also called the {\em double layer potential}).\index{double layer potential} The representation of a
Stokes flow in terms of a single-layer potential is possible, if there is no
net flow through the surface of the capsule \cite{Pozrikidis1992},
$
 \int \! \mathrm{d}A\, (\vec{u}-\vec{u}^\infty)\cdot \vec{n} = 0 \text{,}
$
which is the case for the no-slip boundary condition we are interested in.

In the case of axisymmetry we can integrate over the polar angle and find the
general solution\footnote{The elements of the matrix kernel $\mathsf{M}$ can
  be expressed in terms of elliptic integrals, see for example
  Ref.\ \citenum{Pozrikidis1992}.} 
of the Stokes equation for an axisymmetric
point force distribution,
\begin{align}
u_\alpha (\vec x) - u_\alpha^{\infty}(\vec{x}) &= - \frac{1}{8\pi\mu} \int_C
\mathrm{d}s(\vec{y}) \mathsf{M}_{\alpha\beta}(\vec{y},\vec{x}) f_\beta(\vec{y})
\text{.}
\label{eq:sol1}
\end{align} 
Here, Greek indices denote the components in cylindrical coordinates, i.e.,
$\alpha,\beta= r, z$ ($u_\phi=f_\phi=0$ for symmetry reasons).
The integration in \eqref{eq:sol1} runs along the path $C$ given by the 
generatrix, i.e., the cross section of the 
 boundary $\partial \mathfrak{B}$, with arc length $s(\vec{x})$.

According to the ``no-slip'' condition this results in the equation
\begin{align}
 u_\alpha^\infty  &= \frac{1}{8\pi\mu} \int_C
\mathrm{d}s(\vec{y}) \mathsf{M}_{\alpha\beta}(\vec{y},\vec{x})
f_\beta(\vec{y})
&&(\text{for}~\vec x \in \partial \mathfrak{B}) \text{.}
\label{eq:sol2}
\end{align}

To numerically solve the integral equation for the surface force
$f(\vec{x}_i)$ at a given set of points $\{\vec{x}_i\}$ ($i=1,\ldots,N$) one
can employ a collocation method, the most simple case of which is to choose a
discretised representation of the function $f_\beta(\vec{x})$ and approximate
the integral in \eqref{eq:sol2} by the rectangle method leading to a system of
linear equations. We note that there are (integrable) logarithmic
singularities in the diagonal components of $\mathsf{M}$ which have to be
taken care of.

We can restrict our computations to the bare minimum, i.e., the
surface forces needed for the calculation of the capsule shape but, thereby,
have all necessary information to reconstruct the whole velocity field in the
surrounding liquid. The possibility to limit the computation to the needed
surface forces is one advantage of this approach to the solution of the
Stokes equation in comparison to other approaches that rely on the velocities
or the stream function in the whole domain \cite{HappelBrenner1983,langtangen}.

We assumed a ``no-slip''-condition, that is the velocity directly at the
surface of the immersed body vanishes in its resting frame. This is easily
extended to the case of a non-vanishing tangential slip velocity\footnote{A
  normal velocity on the surface in the capsule's resting frame would conflict
  with its impenetrability and could also lead to a net flux of fluid through
  the capsule, which we cannot incorporate using only the single-layer
  potential.}  is possible, however, to extend this boundary integral approach
to incorporate a prescribed velocity field on the surface in the resting frame
of the capsule \cite{Pozrikidis1992}.  This will allow us to generalise the
approach to model {\it active} swimmers \cite{Lauga2009,degen2014} whose
active locomotion can be captured by means of an effective flow field which is
called the {\em squirmer model} \cite{lighthill, blake1971}.

\section{Iterative solution of shape, flow and sedimenting  velocity}

\begin{SCfigure}[50][ht]
\vspace{-0.3em}
 \includegraphics[width=0.55\linewidth]{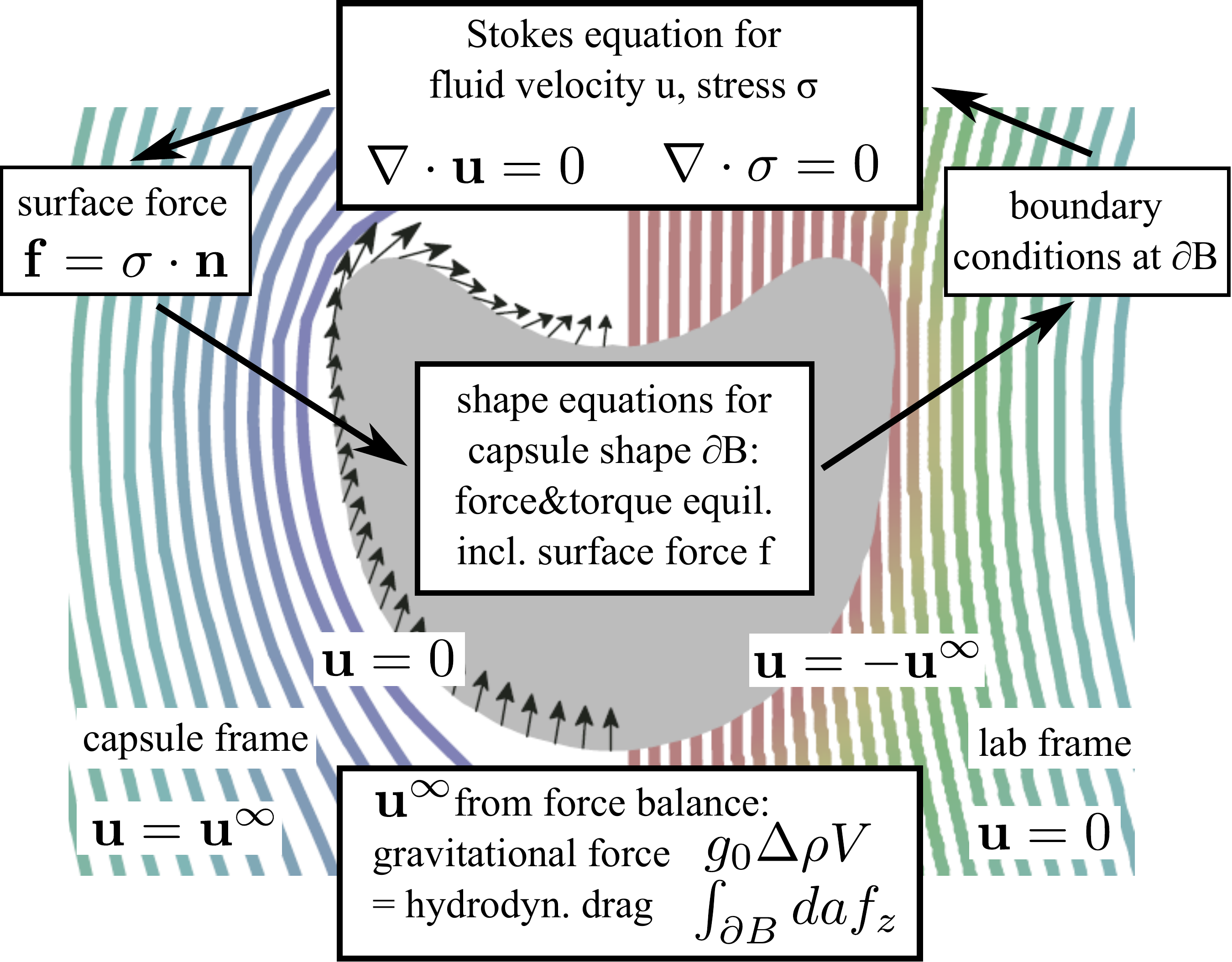}
 \caption{
  Iterative scheme for the solution to the problem of
elastic capsules in Stokes flow allowing for the separation 
of the joint problem into two simpler static problems. 
Details of the iterative scheme are given in
the main text.}
\vspace{-0.3em}
 \label{fig:feedback}
\end{SCfigure}

We find a joint solution to the shape equations and
the Stokes equation by solving them separately and iteratively,
as illustrated in the scheme in  Fig.\ \ref{fig:feedback},
to converge to the desired solution:
We assume a fixed axisymmetric shape and calculate the resulting 
hydrodynamic forces on the capsule  for this shape. 
Then, we use the resulting  hydrodynamic  surface force density 
to calculate a new deformed shape. 
Using this new shape we re-calculate the hydrodynamic surface forces
and so on. 
We iterate until a fixed point is reached. 
At the fixed point, our approach is self-consistent, i.e., the 
capsule shape from which hydrodynamic surface forces are 
calculated is identical to  the capsule shape that is 
obtained by integration of the shape equation under the influence
of exactly these  hydrodynamic surface forces. 

For each capsule shape during the iteration, we can determine 
its  sedimenting velocity $u= |\vec{u}^\infty|$ by 
requiring that the total hydrodynamic drag force 
equals the total driving force.
Because the 
 Stokes equation is linear in the velocity, the force equality is
achieved by just rescaling
 the resulting surface drag forces accordingly via  changing  the
velocity parameter $u$. The velocity therefore plays a similar role 
as a Lagrange multiplier for global force balance.   
In this way, the global force balance can be treated the
same way as other possible constraints like a fixed volume. 
Numerically, it is
impossible to ensure the exact equality of the drag
 and the drive forces, that is $X(L)\equiv 0$ (see eq.\ \eqref{eq:appX}). 
Demanding a very small residual force difference 
 makes it difficult to find an adequate
velocity, a too large force difference
 makes it impossible to find a solution with
small errors at the matching points.

The iteration starts with a given (arbitrary) stress, e.g.,
 one corresponding to
the flow around the reference shape. For the resulting initial 
capsule shape,  the Stokes flow is
computed and the resulting stress is then used to start the iteration. 
If, during the iteration,  the new and old stress
differ strongly it might be difficult to find the new shooting parameters 
for the capsule shape and
the right sedimenting velocity starting at their old values. 
To overcome this technical problem, one can use a convex
combination $\sigma = \alpha \sigma_{\text{new}} + (1-\alpha)
\sigma_{\text{old}}$ of the two stresses and slowly increase the contribution
$\alpha$ of
the new stress until it reaches unity. 
The resulting  capsule shape for $\alpha=1$ is used to continue
the iteration. 
The iteration continues until the change within one iteration 
cycle is sufficiently
small. If there are multiple stationary solutions at a given gravitational
strength the iterative procedure 
 will obviously only find one. Therefore one has to use
continuation of solutions to other parameters (different driving strength
or bending modulus) and possibly multiple initial flows 
to get closer to the full set of solutions.

\section{Application: Sedimenting Hookean capsules} \index{Hookean capsule}

As an application of the outlined method that illustrates the interplay of
elasticity and low Reynolds-number hydrodynamics we consider the {\em
  sedimentation} of Hookean capsules. 
 Sedimentation refers to the motion
under the influence of gravity. However, an effective (and several orders of
magnitude stronger) homogeneous body force 
 can be created within a centrifuge. Thus, on a more
general level we consider an external stress field of the form
$p_{\text{ext}} = - g_0 \Delta \rho z$.
Here, $g_0$ is the gravitational acceleration and $\Delta \rho=\rho_{\rm
  in}-\rho_{\rm out}$ the density difference between the fluids inside and
outside the capsule. Note that we measure the gravitational hydrostatic
pressure $- g_0 \Delta \rho z$ relative to the lower apex, for which we chose
$z(0)=0$. We consider a capsule filled with an incompressible liquid and we
therefore determine the static pressure imposing a volume constraint.

Non-dimensionalising this system using the capsule's equilibrium radius $R_0$ 
and its elastic  modulus $Y_{\rm 2D}$, 
the remaining free parameter are the strength
of the gravitational pull (the {\em Bond number} ${\mathrm Bo} \equiv 
{g_0 \Delta \rho R_0^2}/{Y_{\rm 2D}}$)
and the bending energy relative to the stretching
energy (the inverse 
{\em F{\"o}ppl-von-K\'{a}rm\'{a}n number} $1/\gamma_{\rm FvK}\equiv  
  {E_B}/{Y_{\rm 2D} R_0^2}$).

\begin{figure}[ht]
\centering
 \includegraphics[width=0.99\linewidth]{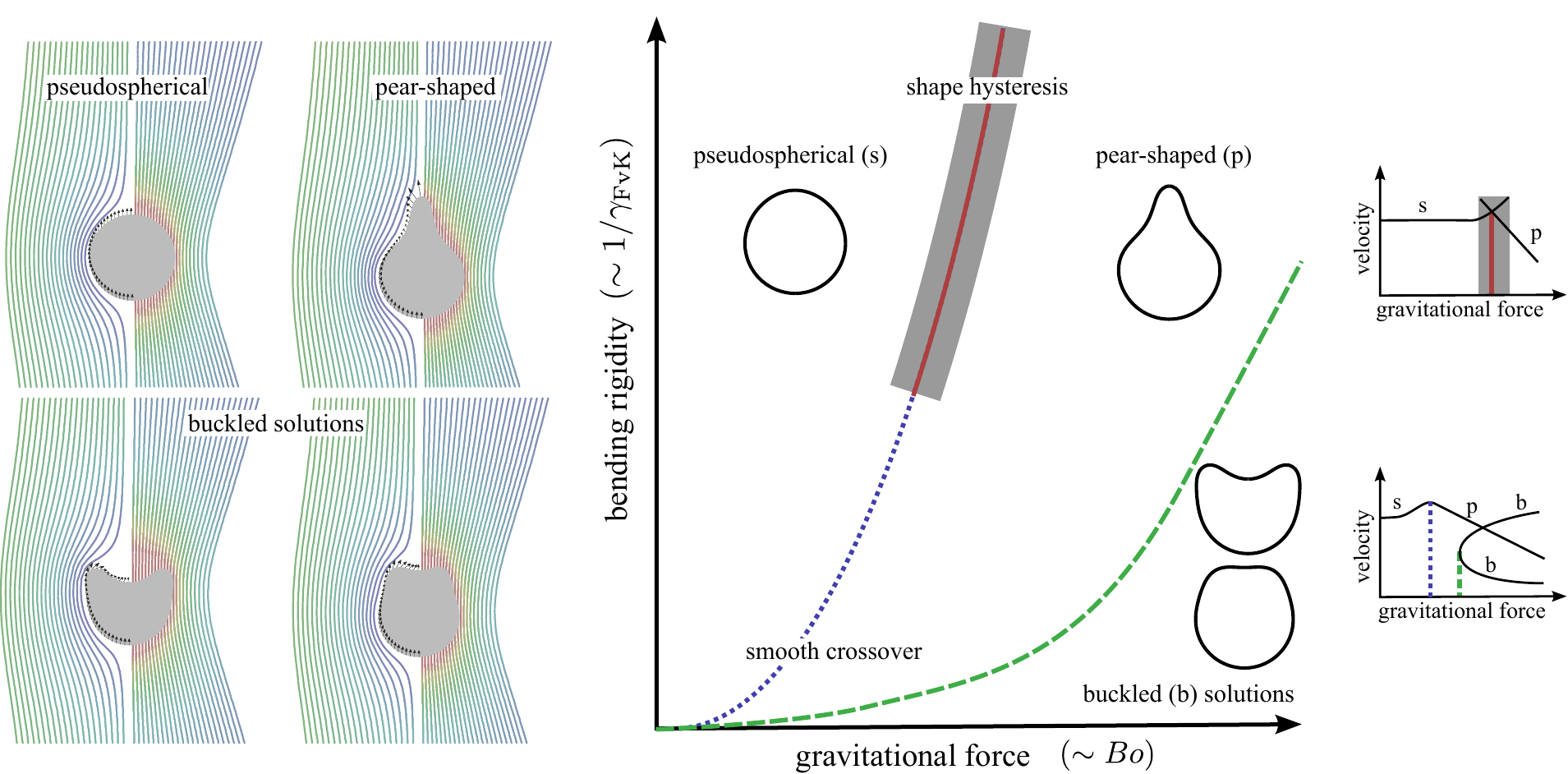}
 \caption{
 Schematic shape diagram of a sedimenting elastic capsule. The free
   parameters of the system are dimensionless 
   bending rigidity and  gravitational pull.
 We find four stable shapes, which are shown on the left 
together with the corresponding fluid streamlines (left: capsule frame, right:
lab frame) and surface forces. 
   The shaded area in the shape diagram in the middle 
  indicates the area of coexistence of the
   pseudospherical and pearshaped solution branches, the blue dotted 
  line is a crossover line. Along the green dashed line the two 
 buckled shapes occur in a bifurcation.
   On the right 
  we  show schematics of
   typical force-velocity relations (depicting the ratio of the velocity of the
   capsule and the velocity of a rigid sphere of same volume as a function of
   the pull) for the two cases of low bending rigidities 
   (no shape hysteresis with the additional branch of
   buckled solutions) and high bending rigidities 
  (with shape hysteresis).
}
 \label{fig:pd}
 \end{figure}

A schematic diagram 
of the stationary shapes that are 
found in this two-dimensional control parameter  space is shown in
Fig.\ \ref{fig:pd}. From Stokes' solution for the flow around a sphere we see
that the acting surface forces are
 $\sigma\cdot\vec{n} = \nicefrac{3}{2 R}\, \mu u
\vec{e}_z$. Thus, the viscous drag tends to stretch the capsule. Additionally
the hydrostatic (gravitational) pressure effectively acts extensionally on the
lower apex and compressional on the upper apex. This leads to a deformation
towards pear shapes. For high bending rigidities the 
indentation of the flanks of
the conical extension is suppressed by bending moments, leading to direct
transition including shape hysteresis.

As in the static problem, the capsule can release stretching stress through
buckling at sufficiently high external stress. However, in this problem these
buckled solutions are {\em coexistent} with the pseudospherical/pear-shaped
solutions.

\section{Conclusion and Outlook}

We showed that the joint problem of an elastic capsule's motion in a viscous
liquid at low Reynolds numbers 
can be reduced to iteratively solving two essentially static
sub-problems, the elastic shape problem for fixed hydrodynamic forces 
and the stationary hydrodynamic Stokes flow problem for fixed boundary
conditions from the capsule
shape,    if one is only interested in the stationary solution.
We  derived the relevant equations of motion and showed how to solve
each the shape of an elastic capsule under a static external hydrodynamic 
stress  and the flow
field of a viscous liquid at low Reynolds number around a rigid
(axisymmetric) body. We then combined these two sub-problem solutions
 and closed the problem by
demanding stationarity under iteration.

Using this iterative method, we are able to resolve coexisting
branches of stationary solutions in the problem of 
``passively'' sedimenting  elastic capsules.
The method can also  be  adapted to ``actively''  swimming deformable 
objects. 
The most direct adaptation is possible for a deformable squirmer,
where an elastic capsule with spherical 
rest shape generates a finite tangent slip velocity.  This will  change 
the boundary conditions of the viscous flow at the capsule surface 
from a no-slip boundary to a given tangential slip velocity 
(in the capsule frame). 
It is also conceivable to treat even  more complex problems, such as 
a diffusiophoretic deformable swimmer, where one eventually  has to 
include the solution of  an appropriate diffusion equation as a 
third coupled sub-problem into the iterative procedure.

\end{document}

%% file: sketch_slab.pdf_tex
\begingroup%
  \makeatletter%
  \providecommand\color[2][]{%
    \errmessage{(Inkscape) Color is used for the text in Inkscape, but the package 'color.sty' is not loaded}%
    \renewcommand\color[2][]{}%
  }%
  \providecommand\transparent[1]{%
    \errmessage{(Inkscape) Transparency is used (non-zero) for the text in Inkscape, but the package 'transparent.sty' is not loaded}%
    \renewcommand\transparent[1]{}%
  }%
  \providecommand\rotatebox[2]{#2}%
  \ifx\svgwidth\undefined%
    \setlength{\unitlength}{152.22463379bp}%
    \ifx\svgscale\undefined%
      \relax%
    \else%
      \setlength{\unitlength}{\unitlength * \real{\svgscale}}%
    \fi%
  \else%
    \setlength{\unitlength}{\svgwidth}%
  \fi%
  \global\let\svgwidth\undefined%
  \global\let\svgscale\undefined%
  \makeatother%
  \begin{picture}(1,0.97288016)%
    \put(0,0){\includegraphics[width=\unitlength]{sketch_slab.pdf}}%
    \put(0.72004092,0.2364624){\color[rgb]{0,0,0}\makebox(0,0)[lb]{\smash{$\tau_\phi$}}}%
    \put(0.50199143,0.76909729){\color[rgb]{0,0,0}\makebox(0,0)[lb]{\smash{$q$}}}%
    \put(0.15864164,0.83587788){\color[rgb]{0,0,0}\makebox(0,0)[lb]{\smash{$m_s$}}}%
    \put(0.7273778,0.58807582){\color[rgb]{0,0,0}\makebox(0,0)[lb]{\smash{$m_\phi$}}}%
    \put(0.36925832,0.85625871){\color[rgb]{0,0,0}\makebox(0,0)[lb]{\smash{$\tau_s$}}}%
  \end{picture}%
\endgroup%